\documentclass[10pt,conference]{IEEEtran}
\IEEEoverridecommandlockouts

\usepackage{cite}
\usepackage{amsmath,amssymb,amsfonts}
\usepackage{algorithmic}
\usepackage{graphicx}
\usepackage{textcomp}
\usepackage{xcolor}
\usepackage[hidelinks]{hyperref}
\usepackage{tabularx}
\usepackage{booktabs}
\usepackage{comment}
\usepackage{enumitem}
\usepackage[linesnumbered,ruled,vlined,noend]{algorithm2e}
\usepackage[skins]{tcolorbox}
\usepackage{subfigure}
\usepackage{listings}
\usepackage{multirow}
\usepackage{array}
\usepackage{makecell}
\usepackage{float}
\usepackage{threeparttable}
\usepackage{adjustbox}
\def\BibTeX{{\rm B\kern-.05em{\sc i\kern-.025em b}\kern-.08em
    T\kern-.1667em\lower.7ex\hbox{E}\kern-.125emX}}

\lstnewenvironment{Report}[1][]{
  \lstset{
    basicstyle =\fontsize{8}{8}\selectfont\ttfamily,
    frame      = tb,
    mathescape = true
  }
}{}

\SetKwRepeat{Do}{do}{while}

\makeatletter
\patchcmd\algocf@Vline{\vrule}{\vrule \kern-0.4pt}{}{}
\patchcmd\algocf@Vsline{\vrule}{\vrule \kern-0.4pt}{}{}
\makeatother

\SetCommentSty{mycommfont}
\SetKwComment{Comment}{\ \#\ }{}
\SetKwProg{Func}{function}{}{end}

\newcommand{\Code}[1]{\begin{small}\texttt{#1}\end{small}}
\newcommand{\SmallCode}[1]{\begin{scriptsize}\texttt{#1}\end{scriptsize}}

\newcommand{\BenchmarkName}{\textsc{LiCoEval}}

\tcbset{
  my box2/.style={
    enhanced,
    colframe=#1!80,
    colback=#1!5,
  },
}
\newtcolorbox{summary-rq}{
  my box2=black,
  boxrule=1pt,top=3pt,bottom=3pt,left=4pt,right=4pt
}
    
\begin{document}

\title{LiCoEval: Evaluating LLMs on License Compliance in Code Generation}

\author{\IEEEauthorblockN{Weiwei Xu\IEEEauthorrefmark{1}, Kai Gao\IEEEauthorrefmark{4}, Hao He\IEEEauthorrefmark{2}, Minghui Zhou\thanks{\IEEEauthorrefmark{3}Minghui Zhou is the corresponding author.}\IEEEauthorrefmark{1}\IEEEauthorrefmark{3}}
\IEEEauthorblockA{\IEEEauthorrefmark{1}\textit{School of Computer Science, Peking University, Beijing, China}\\
\IEEEauthorrefmark{1}\textit{Key Laboratory of High Confidence Software Technologies, Ministry of Education, China}\\
\IEEEauthorrefmark{4}\textit{University of Science and Technology Beijing, Beijing, China}\\
\IEEEauthorrefmark{2} \textit{Carnegie Mellon University, Pittsburgh, USA}\\
xuww@stu.pku.edu.cn, kai.gao@ustb.edu.cn, haohe@andrew.cmu.edu, zhmh@pku.edu.cn}
}
\maketitle

% Force page numbers, make sure to remove in camera ready
\thispagestyle{plain}
\pagestyle{plain}
\begin{abstract}
Recent advances in Large Language Models (LLMs) have revolutionized code generation, leading to widespread adoption of AI coding tools by developers. 
However, LLMs can generate license-protected code without providing the necessary license information, leading to potential intellectual property violations during software production. 
This paper addresses the critical, yet underexplored, issue of license compliance in LLM-generated code by establishing a benchmark to evaluate the ability of LLMs to provide accurate license information for their generated code.
To establish this benchmark, we conduct an empirical study to identify a reasonable standard for ``striking similarity" that excludes the possibility of independent creation, indicating a copy relationship between the LLM output and certain open-source code. 
Based on this standard, we propose \BenchmarkName,
to evaluate the license compliance capabilities of LLMs, i.e., the ability to provide accurate license or copyright
information when they generate code with striking similarity to already existing copyrighted code. 
Using \BenchmarkName, we evaluate 14 popular LLMs, finding that even top-performing LLMs produce a non-negligible proportion (0.88\% to 2.01\%) of code strikingly similar to existing open-source implementations. 
Notably, most LLMs fail to provide accurate license information, particularly for code under copyleft licenses.
These findings underscore the urgent need to enhance LLM compliance capabilities in code generation tasks. 
Our study provides a foundation for future research and development to improve license compliance in AI-assisted software development, contributing to both the protection of open-source software copyrights and the mitigation of legal risks for LLM users.
\end{abstract}

\section{Introduction}
Recent advances in Large Language Models (LLMs) have instigated revolutionary changes in the fields of artificial intelligence, natural language processing, and software engineering~\cite{dong2024generalization}.  
With billions of parameters trained on extensive corpora (both general and software engineering specific), LLMs have demonstrated extraordinary competencies in various software engineering tasks, such as code generation~\cite{fried2022incoder,li2023starcoder, izadi2022codefill}, program repair~\cite{jin2023inferfix,xia2023automated,wei2023copiloting}, and documentation generation~\cite{luo2024repoagent}.
Specifically, LLMs' remarkable code generation capabilities enabled the rapid adoption of AI coding tools in practice.
As GitHub reports, 92\% of US-based developers are already using AI coding tools both inside and outside of work~\cite{githubsurvey}. 

However, the widespread utilization of LLMs for code generation has also elicited concerns regarding security~\cite{pearce2022asleep, perry2023users}, privacy~\cite{yang2024unveiling, Al2024traces}, and legal issues~\cite{al2023ab, choksi2023whose, henderson2023foundation, yu2023codeipprompt}. A key issue among these is the potential infringement of intellectual property (IP) rights of a vast number of open-source developers~\cite{ yu2023codeipprompt, yang2024unveiling, karamolegkou2023copyright}. Due to the fact that LLMs are trained on extensive volumes of open-source code governed by open-source licenses and their inherent ability to recognize and memorize patterns~\cite{yang2024unveiling}, they may generate code snippets that are similar or even identical to those in the training data under certain prompts~\cite{yang2024unveiling, Al2024traces,yu2023codeipprompt}. 
% \begin{figure}
%     \centering
%     \includegraphics[width=\linewidth]{Figs/similar_example.pdf}
% \vspace{-6mm}
%     \caption{A real instance on Twitter, complained by a developer, where Copilot generated code that closely resembled his open-source code but did not include the appropriate copyright information~\cite{TimTwitter}.}
% \vspace{-4mm}
%     \label{fig:similarexample}
% \end{figure}
Consequently, the use of these generated code must comply with the terms and conditions in the open-source license, concerning how a piece of open-source software (OSS) can be reused, modified, and redistributed~\cite{xu2023licensecom,german2009license}. %%i thought all three xu are weiwei, apparently that's the author of this paper to a reviewer
%% xu2023lidetector is not weiwei's paper
Licenses, which are unavoidably linked to the open-source code, serve a dual purpose: they not only enforce the developers' commitment to sharing, transparency, and openness~\cite{Al2024traces, al2023ab}, but also necessitate that those who reuse the code respect and adhere to the terms set by the original authors~\cite{wolter2023open,huang2023detecting}. 
For example, under \emph{Apache 2.0}~\cite{Apache}, authors grant users perpetual copyright and patent rights, but users must also comply with the corresponding license terms such as including attribution notices and modification statements when redistributing the software. 
\begin{figure*}
    \centering
    \includegraphics[width=\linewidth]{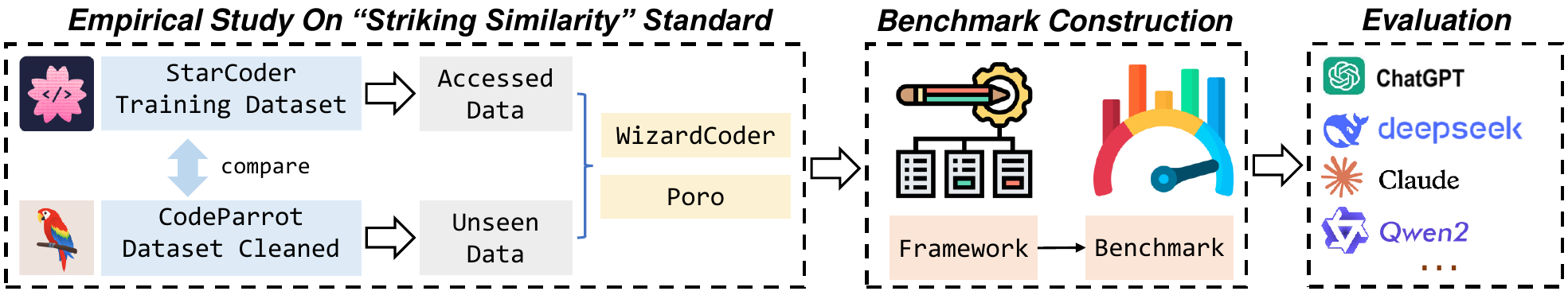}
\vspace{-3mm}
    \caption{Overview of this study.}
\vspace{-3mm}
    \label{fig:overview}
\end{figure*}

Despite their extraordinary competencies, LLMs often overlook license information when they memorize and generate code identical to the works of open-source developers. 
This has sparked a growing wave of concern and discontent among the open-source community~\cite{StefanTwitter, TimTwitter, ChrisTwitter,eeveeTwitter}. For example, an open-source developer accuses Copilot of \emph{``emitting large chunks of my copyrighted code, with no attribution, no LGPL license."}~\cite{TimTwitter}
% As shown in Figure~\ref{fig:similarexample}, these two pieces of code come from an open-source developer who, on social media, accuses Copilot of ``emitting large chunks of my copyrighted code, with no attribution, no LGPL license"~\cite{TimTwitter}. 
More than just ethical concerns, these actions by LLMs may also pose unpredictable legal risks to users. 
For example, if LLMs produce open-source code without providing the necessary license information, users are naturally unable to utilize such code in legal compliance processes~\cite{yu2023codeipprompt, yang2024unveiling}. 
It should be noted that OpenAI, GitHub, and Microsoft are currently facing lawsuits, as GitHub Copilot~\cite{GithubCopilot} has been implicated in reproducing licensed code without compliance with the corresponding license terms~\cite{Copilotlitigation}.
These controversies underscore that the license compliance capabilities of LLMs, i.e., \textbf{\textit{the ability to provide accurate license and copyright information during code generation}}, play a crucial role in both protecting the IP rights of numerous open-source developers and shielding users of such models from unforeseen legal risks. Therefore, it is important to evaluate the license compliance capability of LLMs for code generation tasks.

% With most LLMs' training data being undisclosed, it becomes challenging to ascertain whether the model has \emph{accessed and memorized} the open-source code for its output, or whether its capability is so advanced that it can produce code similar to that of human authorship, even without prior exposure to the specific code. According to the principle, the former scenario requires license compliance, while the latter does not. 

%% need details
To the best of our knowledge, it remains unclear how well LLMs perform in terms of license compliance despite the substantial effort devoted to evaluating LLMs' performance on code completion accuracy~\cite{chen2021evaluating,du2023classeval,yu2024codereval}, robustness~\cite{yang2022natural,pour2021search}, security~\cite{pearce2022asleep,aghakhani2023trojanpuzzle}, and privacy~\cite{yang2024unveiling,niu2023codexleaks}, among others.
To fill this knowledge gap, this paper makes the first attempt to evaluate the license compliance capabilities of LLMs by establishing a novel benchmark. 
This evaluation of LLMs encompasses considerations of not only a technical nature but also legal aspects.
Specifically, \textbf{one of the most challenging aspects of assessing the compliance capabilities of LLMs is determining whether their outputs, when similar to specific open-source code snippets, are derived from those snippets or are independently created yet merely coincidentally similar.} 
The former scenario would entail issues of license compliance, whereas the latter does not.
% Indeed, one of the most challenging aspects of assessing the compliance capabilities of LLMs is determining whether their output, similar to open-source code, exists copying. 
% This determination is crucial because it helps distinguish between potentially infringing outputs and those that are independently generated. 
A key and universally acknowledged principle used in law to aid in this distinction is ``\emph{access and substantial similarity}"~\cite{USCourt, scheffler2022formalizing}, %%no need to have many, but keep the most basic and important ones
which means that potential infringement only occurs when a party has \emph{access} to a work and the resulting product is \emph{substantially similar} to that work. 
Alternatively, in the absence of evidence indicating direct access, the existence of copying can be established when the degree of similarity between two works is so striking as to preclude the possibility of independently arriving at identical outcomes~\cite{USCourt,Arnsteinv.Porter,autry2002toward,higgins2003proving}.
Considering that most LLMs' training data are undisclosed, it is challenging to determine and verify whether an LLM has accessed a specific code snippet. Therefore, we focus on scenarios where the outputs of LLMs are \emph{strikingly similar} to open-source code, to assess their compliance capabilities in this context.

As demonstrated in Figure~\ref{fig:overview}, to establish the benchmark, we first conduct an empirical study to explore a reasonable standard of \emph{striking similarity}, to distinguish between cases where LLMs generate code without prior exposure to similar open-source instances and cases where LLMs memorize open-source code in the training data.  
%This endeavor is primarily motivated by the complex nature of intellectual property infringement cases, which are typically adjudicated on a case-by-case basis in the absence of a unified standard~\cite{scheffler2022formalizing}. Furthermore, the case of LLMs independently generating similar code without prior exposure to specific instances remains an area of legal vacuum. 
Based on the findings of this empirical study, we propose an evaluation benchmark, \BenchmarkName,~for LLMs' \underline{li}cense \underline{co}mpliance capability \underline{eval}uation, and then evaluate 14 popular LLMs currently in use for code generation. 
We find even top-performing code generation LLMs produce a non-negligible proportion (0.88\% to 2.01\%) of strikingly similar output compared to existing open-source code. Most models fail to provide any license information for code snippets under copyleft licenses, with only \Code{Claude-3.5-sonnet} demonstrating some ability to correctly provide license information for such code. These results highlight the urgent need to improve license compliance capabilities in LLMs used for code generation, particularly in handling code under copyleft licenses and providing accurate license information.

In summary, the contributions of this paper are as follows:
\begin{itemize} 
    \item We conduct an empirical study to explore a preliminary standard for ``\emph{striking similarity}" under the legal context of intellectual property infringement and LLM's capabilities in independently generating non-trivial code.
    \item We design a framework for evaluating the license compliance capabilities of LLMs in code generation and provide the first benchmark for evaluating this capability.
    \item We evaluate the license compliance capabilities of 14 popular LLMs, providing insight into improving the LLM training process and regulating LLM usage.
\end{itemize}

\section{Background and Related work} %below you very clearly state (several times) the rational for exploring standard of strikingly similarity, and mentioned evaluation, but where is the evaluation framework (it's ok if it is not the point)?
\subsection{License Compliance and IP Infringement}
\label{sec:LiCom}
Open-source software authors grant copyrights and, in some cases, patent rights within IP right through open-source licenses. However, this grant is conditional and the licenses specify the obligations to which users must comply~\cite{Meeker2020OpenSource, rosen2005open,wu2024large}. 
It is important to note that open-source authors do not oppose or prohibit the reuse of their code; in fact, such reuse aligns precisely with their foundational motives for choosing to open-source their code. 
Nevertheless, they require users to comply with the stipulations set forth in the licenses when reusing the code. Failure to do so can lead to license non-compliance, ultimately resulting in intellectual property infringements~\cite{ViolatingLicense}. 

In the context of LLM code generation, if models output licensed open-source code without providing license information, users reusing this code might face compliance risks. 
This concern is particularly relevant for companies and organizations, as it could lead to copyright infringement liabilities and potentially result in significant economic compensation and harm to their reputation and operations~\cite{xu2023licenserec}.

In the judicial context, it is crucial to determine whether the content generated by LLMs has a copying relationship with specific open-source code. According to existing legal principles, copying can be established when similarities between two works are so striking that they preclude the possibility of independently arriving at the same result~\cite{USCourt,Arnsteinv.Porter}. 
However, the definition of ``\emph{striking similarity}" remains ambiguous. Courts consider several factors in their analysis, such as the uniqueness, intricacy, or complexity of similar sections, and the appearance of the same errors or mistakes in both works~\cite{autry2002toward}. The ambiguous definition, coupled with the impressive capabilities of LLMs in independently generating non-trivial code snippets, motivates our empirical study to explore reasonable standards for identifying \emph{striking similarity} in the context of LLM generated code.

%%there is a bunch of code clone techniques that need to be mentioned as related work somewhere

\begin{comment}%%
It is crucial to acknowledge that in the judicial system, particularly when dealing with intellectual property infringement cases, each case is individually assessed based on its unique circumstances~\cite{rafiqi2013copyright}. Our research aims to explore standards that can reflect the risks associated with LLMs to a certain extent. However, this does not imply the establishment of actual legal boundaries. 
The findings of our study are intended to serve as a guide for understanding potential issues, rather than providing conclusive legal judgments.
\end{comment}
% In the judicial context, a generally recognized principle for assessing intellectual property infringement is "\emph{access and substantial similarity}"~\cite{USCourt,USCopyrightLaw, scheffler2022formalizing,balganesh2014judging}.

% This principle serves as a critical framework for determining whether an infringement has occurred. It posits that infringement may be established if it can be shown that the alleged infringer had access to the original work and that the work they produced bears a substantial similarity to the copyrighted material. The concept of "\emph{access}" ensures that the similarity is not coincidental, establishing a direct or indirect connection between the original work and the alleged infringing work. Therefore, in our work, judging \emph{access} is a pivotal aspect for evaluating LLMs on license compliance. 
 
\subsection{Memorization in LLMs for Code}
\label{sec:MemInLLMs}
Research consistently demonstrates that general LLMs tend to memorize content from their training sets,  especially in larger models~\cite{zhang2021counterfactual,carlini2022quantifying}, raising significant concerns regarding IP rights violations~\cite{carlini2022quantifying,karamolegkou2023copyright,lee-etal-2022-deduplicating}. This memorization also occurs in LLMs for code, potentially causing compliance issues through unintentional output of licensed code  
~\cite{yang2024unveiling,yu2023codeipprompt,Al2024traces}.
Recent studies on memorization~\cite{yang2024unveiling,Al2024traces} and IP infringement~\cite{yu2023codeipprompt} in LLMs for code, %% is there a term called "code LLM"?
particularly those trained on non-public datasets, typically compare LLM outputs with existing open-source code to detect memorization.
% For instance, Al-Kaswan et al.~\cite{Al2024traces} uses exact match rate (EM) and BLEU-4 score, and Yu et al.~\cite{yu2023codeipprompt} employs code plagiarism detection tools like Dolos~\cite{maertens2022dolos} to calculate the similarity.
% While it has been confirmed that LLMs are likely to output similar code via memorization, it remains uncertain whether they can also accurately provide the corresponding copyright and license information for the code. This uncertainty motivates our design of a benchmark to evaluate the ability of LLMs in this respect, aiming to reflect the potential compliance risks users may face when using such LLMs.
However, with undisclosed training sets, it is challenging to definitively attribute this similarity to memorization rather than LLMs' extraordinary generalization competencies. This ambiguity extends to numerous code clone detection methods, which describe the degree of similarity between code snippets from various perspectives, such as textual similarity, call graphs, and other structural features~\cite{zakeri2023systematic}. Yet, we still lack a clear understanding of what degree of similarity is sufficient to constitute ``striking similarity"   % can we claim "in law"? drop "in law" or change a phrase, we simply borrow this term from law terms
that can exclude the possibility of independent creation.
%Furthermore, it remains uncertain whether LLMs can accurately provide the corresponding copyright and license information for code they appear to have memorized. 

Furthermore, it remains uncertain whether LLMs can accurately provide the corresponding copyright and license information for code they appear to have memorized. 
%The most relevant work~\cite{yu2023codeipprompt} does not address this critical issue, leaving a critical gap in current research. While Yu et al.\cite{yu2023codeipprompt} examine code similarity, there remains a substantial distinction between code similarity and actual IP infringement. Open-source code, by definition, permits reuse under specific conditions; mere reuse does not inherently constitute infringement. The key issue lies in whether the reuse complies with the specified license terms. 
The most relevant work by Yu et al.~\cite{yu2023codeipprompt} investigates to what extent LLMs generate licensed code. However, their study was conducted under a problematic assumption that LLMs should not generate licensed code at all, which fundamentally misunderstands the nature of open-source software. Open-source code, by definition, permits reuse under specific conditions; mere reuse does not inherently constitute infringement. The key issue lies in whether the reuse complies with the specified license terms, which, however, is not investigated in their work and requires non-trivial efforts in the LLM context.
To address these challenges and uncertainties, we conduct an empirical study to establish a standard of \textit{striking similarity} and build a benchmark to evaluate LLM's compliance capability.

\subsection{Evaluations of LLMs for Code Generation}
In the realm of code generation with LLMs, numerous benchmarks have been introduced to evaluate these models' capabilities~\cite{chen2021evaluating,austin2021program,du2023classeval,yu2024codereval}. These benchmarks typically focus on generating code snippets from natural language descriptions, employing metrics such as Pass@k~\cite{chen2021evaluating} to assess the accuracy of the generated code. Furthermore, many studies separately investigate the non-functional properties of LLMs for code~\cite{yang2024robustness}, including robustness~\cite{yang2022natural,pour2021search}, security~\cite{pearce2022asleep,aghakhani2023trojanpuzzle}, privacy~\cite{yang2024unveiling,niu2023codexleaks,Al2024traces}, and explainability~\cite{aleithan2021explainable}. However, to our knowledge, there has been no evaluation focusing on the compliance capabilities of these models.%should we drop "in legal context"? we simply provide technique that may apply for legal context, but we don't attempt to provide a standard for legal context (we simply are not able to do that). Carefully think about it. Note your rq in section III is phrased very good.
This lack of evaluation is disadvantageous for users seeking models with lower legal risks and also hinders model developers from making targeted improvements in this critical area during the training process, which motivates our work.

\section{Empirical Study on Standard of Striking Similarity}
\subsection{Research Question}
\textbf{RQ: Where might the reasonable standard of \emph{striking similarity} lie in the context of code generation by LLMs?}

A significant challenge in evaluating the compliance capabilities of LLMs is determining whether there exists a copying relationship between the output of LLMs and existing open-source code. Due to the remarkable code generation abilities of LLMs, it is plausible that they can independently generate code that is similar to a specific open-source code snippet, even without prior exposure to it. If the code is independently generated, it clearly does not necessitate the provision of corresponding copyright information.

In accordance with the legal principle of \emph{striking similarity}~\cite{USCourt,Arnsteinv.Porter}, 
%i.e., a degree of similarity between two pieces of code that is so striking that it is implausible they were produced independently, 
we aim to explore where the reasonable standard of \emph{striking similarity} might lie for LLMs. 
%Meeting the \emph{striking similarity}  standard would imply a copying relationship between the LLM's output and a specific code snippet, necessitating the provision of corresponding copyright information by the LLM.
%, as it would be otherwise impossible to generate such strikingly similar code.  
%In such circumstances, it would be necessary for the LLMs to provide corresponding copyright information. 
The empirical standard explored in this study lays a foundation for our subsequent evaluation framework and benchmark. 
%As described in Section~\ref{sec:LiCom},  
Our exploration strives to align with copyright law principles in identifying a relatively reasonable standard that can, to some extent, evaluate the compliance risks associated with using certain LLMs. Notably, \textit{our findings are not intended to establish definitive legal boundaries}. Instead, they are intended to serve as a guide for understanding and identifying potential compliance issues, offering insights that can inform future research and development in this rapidly evolving field.

\begin{table}[t]

\centering
\footnotesize
    \caption{Model size and performance(pass@1(\%))  on the HumanEval~\cite{chen2021evaluating}
benchmark for models trained on fully open-source datasets.}
\vspace{-1mm}
\setlength{\tabcolsep}{0.8mm}
  \label{tab: modelinfo}
  \renewcommand{\arraystretch}{1.2}
  \begin{tabular}{llccc}
    \toprule
    Type &Model & Size & HumanEval  & Dataset \\
    \midrule
    General
    &Poro-34B-chat~\cite{luukkonen2024poro} & 34B & 37.2  &  The Stack \\
    \midrule
    \multirow{4}{*}{\makecell{Code}}
    &Starcoder2-15B-Instruct~\cite{lozhkov2024starcoder}& 15B &72.6 & The Stack V2\\
   &\textbf{WizardCoder-15B-V1.0}~\cite{luo2023wizardcoder} & 15B  & 52.4 & The Stack\\ 
    &StarCoder~\cite{li2023starcoder} & 15B & 33.6 & The Stack \\ 
    &Codeparrot~\cite{codeparrot} & 1.5B &3.99 & CodeParrot-Clean\\
   
    \bottomrule
    \end{tabular}
\vspace{-2mm}    
\end{table}

%\subsection{Study Subjects}
%\subsubsection{Model for analysis}
\subsection{Selection of LLMs}
We aim to select representative LLMs to explore the \textit{striking similarity } standards, with the following principles: 
\begin{itemize}
    \item High accuracy in code generation tasks, as license compliance has no meaning without accuracy (a model can generate very chaotic, functionally incorrect code that naturally does not resemble existing open-source code).
    \item All training data including the instruction tuning data is public. We need to ascertain the data that the model has been exposed to.
\end{itemize}
% 1. High accuracy in code generation tasks, as license compliance has no meaning without accuracy (a model can generate very chaotic, functionally incorrect code that naturally does not resemble existing open-source code). 2. All training data is public. In this empirical study, we need to ascertain the data that the model has been exposed to. 
% We present the size and performance of several LLMs that are widely studied in recent code generation research~\cite{zhu2024deepseek,team2024codegemma,lozhkov2024starcoder} in Table~\ref{tab: modelinfo}. %through public data~\cite{mouselinos2022simple,WizardLM,llm-coding-evaluation,luukkonen2024poro,roziere2023code}. %need more justification on the selection, e.g., we extensively invesitigated the possible LLMs (through ....) that can be used for study, this is all we can get...
We identify three widely used open-sourced datasets for training LLMs on the Huggingface platform, i.e., \Code{The Stack}~\cite{the-stack}, \Code{The Stack v2}~\cite{the-stack-v2}, and \Code{CodeParrot Dataset Cleaned}~\cite{codeparrot-clean}. Based on these datasets, we find five popular LLMs trained on these datasets as listed in Table~\ref{tab: modelinfo}. 
Among these LLMs, we select \Code{WizardCoder-15B-V1.0} as our research subject for two reasons. First, it shows an acceptable performance with a Pass@1 score of 52.4 on the HumanEval benchmark. Second, it has a manageable size (i.e., 15B parameters) under our hardware constraints, which is comparable to, or even larger than the LLMs studied in previous work~\cite{yang2024unveiling,Al2024traces,yu2023codeipprompt}. 
\Code{WizardCoder}\footnote{In this paper, ``\SmallCode{WizardCoder}" refers to ``\SmallCode{WizardCoder-15B-V1.0}" in all subsequent mentions.} uses an open-source instruction-following dataset (\Code{Code-Alpaca5}\cite{codealpaca}) to fine-tune the \Code{StarCoder} trained on \Code{The Stack} dataset. 
The complete open-source nature of its training datasets facilitates comprehensive analysis and verification. 
Despite \Code{Starcoder2-15B-Instruct}'s higher Pass@1 score on HumanEval, we chose \Code{WizardCoder} for our analysis due to its smaller, more manageable dataset (200B tokens vs. 900B tokens in \Code{Starcoder2-15B-Instruct}'s ). 
%Additionally, among all models with fully open-source training data, \Code{WizardCoder} demonstrates superior performance. Its smaller dataset (200B tokens) compared to \Code{The Stack v2} (900B tokens) makes it more manageable for our analysis. 
%Given these factors, we select \Code{WizardCoder} as our research subject.

% Among the models with public training sets, WizardCoder-33B-V1.1~\cite{luo2023wizardcoder} performs exceptionally well, surpassing GPT-3.5-turbo on HumanEval~\cite{chen2021evaluating} and comparable with GPT-3.5-turbo on MBPP~\cite{austin2021program}.
%WizardCoder-15B-V1.0\cite{luo2023wizardcoder} is based on Starcoder\cite{li2023starcoder} and has undergone fine-tuning using a code instruction-following training set. Crucially, all of WizardCoder-15B-V1.0's training data is open-source. This includes the Starcoder training set (Starcoderdata\cite{starcoderdata}) and the initial 20K instruction-following dataset (Code-Alpaca5\cite{codealpaca}) used for fine-tuning. The complete transparency of WizardCoder-15B-V1.0's training data is essential for our research objectives, as it allows for comprehensive analysis and verification.
%Given this full transparency in training data, along with its performance and model size, we select WizardCoder~\footnote{In this paper, ``WizardCoder" refers to ``WizardCoder-15B-V1.0 in all subsequent mentions.} as our primary research subject.
\begin{figure}[t]
    \centering
    \includegraphics[width=\linewidth]{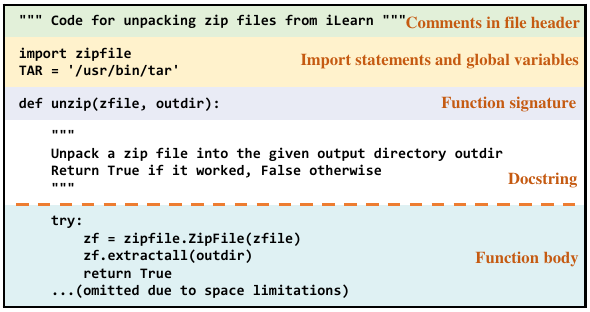}
\vspace{-4mm}
    \caption{Structure of function-level code snippet.}
\vspace{-5mm}
    \label{fig:codeexample}
\end{figure}

%\subsection{Methods}
\subsection{Experiment setup}

%%%%%%%%%need to combine the following paragraph into this section
In order to detect \textit{striking similarity}, we design an experiment as follows.

First, we construct two distinct groups of code samples, \textsc{Unseen} and \textsc{Accessed}, to simulate two different scenarios. 
The first scenario is where the model independently completes the corresponding code, as it has not been exposed to the code from the \textsc{Unseen} group before. 
The second scenario pertains to instances where the generated content cannot be regarded as independently created, potentially implicating a copying relationship. 
This is attributed to the fact that samples within the \textsc{Accessed} group are derived from the model's training dataset. 
We select two groups of code samples for this study as described in Section~\ref{sec:codesamples}.
%%%%%%%%%%%%%%%%%

%To answer this research question, 
Second, we construct prompts using the \textsc{Unseen} and \textsc{Accessed} groups, then instruct \Code{WizardCoder} to complete the code snippets. 
Our goal is to observe potential differences in similarity when the model generates code for these two distinct groups.
As illustrated in  Figure~\ref{fig:codeexample}, we divide the function-level code snippets into five parts: file header comments, import statements and global variables, function signature, docstring, and function body. We combine the first four parts into a prompt with the aim of providing the model with as complete an input context as possible, and then instruct the model generate the function body.
We conduct experiments in a one-shot fashion using greedy decoding (temperature set to 0). %this clearly explains how you conduct your experiment, but you explained Table II in above section when you haven't explained what it means as introduced in this section

Third, we select specific features derived from copyright law principles to characterize \textit{striking similarity} as described in Section~\ref{ss:standardofsimilarity}.
By analyzing the similarity between the LLM's output and the original code snippets when completing tasks from both groups, we aim to identify a relatively reasonable standard for \emph{striking similarity}. 
If the generated code meets this standard for \emph{striking similarity}, it can be inferred that the code may not be an independent creation by the model, indicating a possible copying occurrence.

\begin{figure*}[t]
    \centering
    \includegraphics[width=0.95\linewidth]{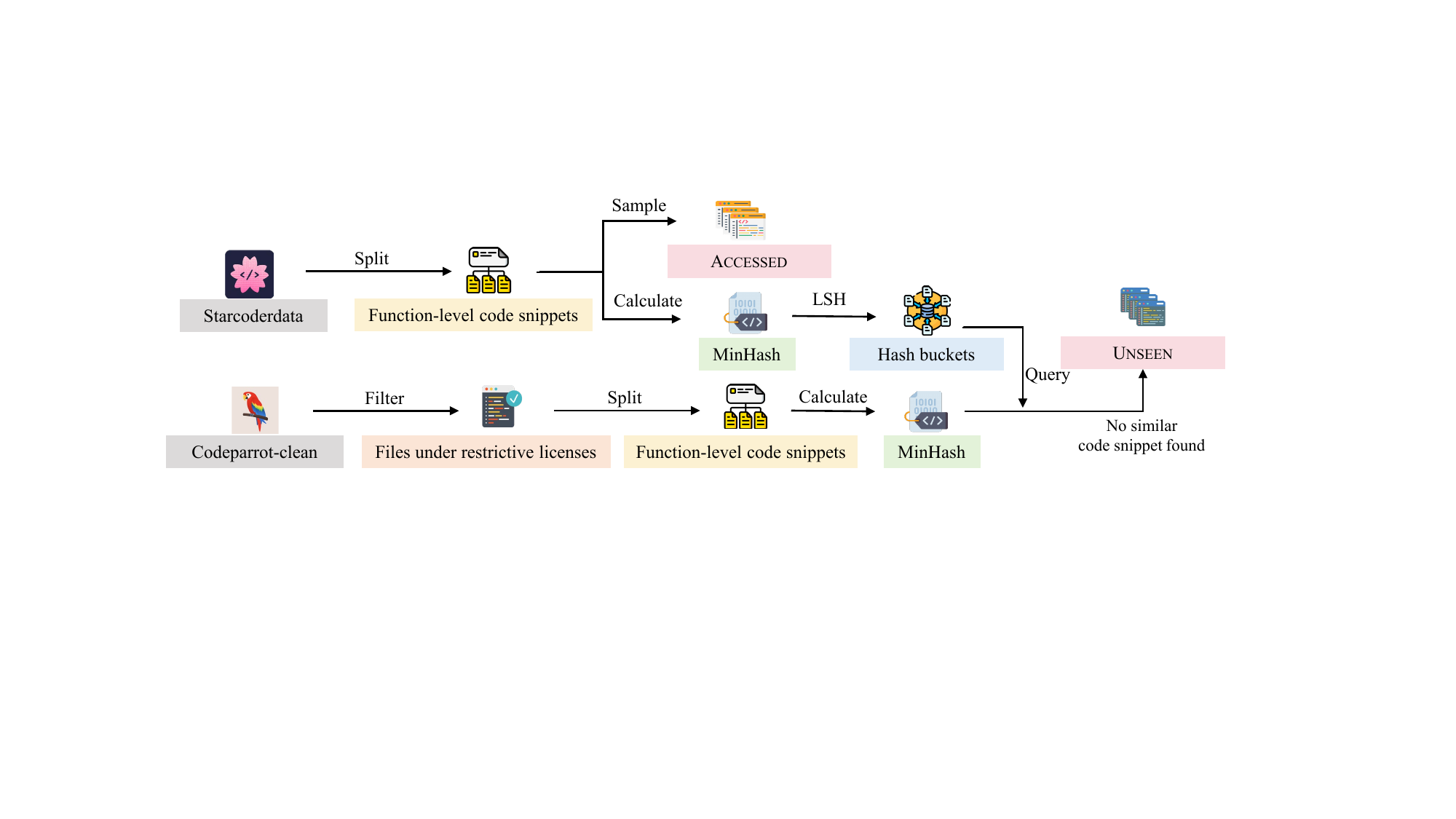}

    \caption{Overview of code samples construction.}
\vspace{-5mm}
    \label{fig:dataoverview}
\end{figure*}

\subsection{Method}
\subsubsection{Construction of Code Samples}
\label{sec:codesamples}
Figure~\ref{fig:dataoverview} illustrates the overview of the construction method. Our code samples originate from two sources: the training set of \Code{WizardCoder}, i.e., \Code{Starcoderdata}\cite{starcoderdata}, and \Code{Codeparrot-clean} dataset~\cite{codeparrot-clean}. 
% WizardCoder is fine-tuned from Starcoder~\cite{luo2023wizardcoder}, with Starcoderdata serving as the training set. 
\Code{Starcoderdata}, a subset of \Code{The stack}, comprises 783GB of code spanning across 86 programming languages. A license filtration mechanism was implemented during the data collection process for \Code{Starcoderdata}, which guarantees that all code files within \Code{Starcoderdata} are sourced from the repositories under permissive licenses~\cite{li2023starcoder}. \Code{Codeparrot-clean} dataset comprises 5,361,373 Python files from GitHub, including those under restrictive licenses such as \textit{GPL-3.0}~\cite{GPLv3}.
The former is utilized to construct the \textsc{Accessed} group that \Code{WizardCoder} has been exposed to, while the latter is used to construct the \textsc{Unseen} group that \Code{WizardCoder} has never encountered before.

Given that LLMs are not yet adept at performing code completion beyond the granularity of functions~\cite{du2023classeval},  we choose to conduct our experiments in alignment with well-known accuracy benchmarks such as HumanEval~\cite{chen2021evaluating}, specifically at the function-level granularity. Moreover, following the existing research on memorization of LLMs for code~\cite{yang2024unveiling,Al2024traces}, our study
focuses on Python considering its prevalence~\cite{srinath2017python}. 

To construct the \textsc{Accessed} group,  we first extract 74,772,489 function-level code snippets from \Code{Starcoderdata}'s Python files. 
% Adhering to the original deduplication strategy of the Starcoderdata dataset, we calculate the MinHash~\cite{srinath2017python} for each function individually. Subsequently, we employ Locally Sensitive Hashing (LSH) to map similar functions into the same bucket. In this progress, we use 5-grams and a Jaccard similarity of 0.5. 
We further select samples based on the following principles, resulting in 2,628,395 function-level code snippets: 1) the function must have a docstring and the function body must be more than six lines (the median value of all functions' lengths), which aims to provide LLMs with more comprehensive descriptions of the function's capabilities and exclude overly trivial code snippets;
% \item The combined length of the file header, function signature, and docstring, i.e., the length of prompt, must not exceed 1024, which is the maximum length allowed during the Starcoder training~\cite{li2023starcoder}.
2) the function should not be a class method due to their typically more complex context dependencies~\cite{du2023classeval};
3) the function-level code snippet has no syntax errors.
Eventually, we randomly sample 10,000 functions 
% (95\% confidence level, 5\% confidence interval~\cite{SampleSizeCalculator}) 
as the \textsc{Accessed} group that \Code{WizardCoder} has been exposed to.

To construct the \textsc{Unseen} group, we select code files from the repositories authorized under restrictive licenses in \Code{Codeparrot-clean} dataset, and segment them into functions. Given that \Code{Starcoderdata} excludes code from such repositories, it is highly probable that \Code{WizardCoder} has not encountered these code snippets in its training set. This makes these functions our preliminary candidate samples. 
To ensure these functions are truly unseen by \Code{WizardCoder}, we need to compare each function with  the entire training set. 
% The goal is to meticulously identify those code snippets that are truly unseen to WizardCoder. 
However, due to the large scale of the dataset, direct comparison of each function with every training sample is impractical. We therefore adopt the deduplication strategy used in \Code{Starcoderdata}~\cite{li2023starcoder}. We calculate a MinHash~\cite{broder2000identifying} for each function in the training set and use Locality-Sensitive Hashing (LSH) to efficiently map similar functions into the same bucket, using 5-grams and a Jaccard similarity threshold of 0.2 (more stringent than the 0.5 threshold used in \Code{Starcoderdata}) for this process.
Then, we compute the MinHash for each function in preliminary candidate samples from \Code{Codeparrot-clean} dataset and query these in the LSH buckets of the training set. Functions without similar matches in the training set are considered final candidates.
We apply the same three principles used in constructing \textsc{Accessed} group and sample 10,000 samples from 157,273 final candidates 
% (95\% confidence level, 5\% confidence interval~\cite{SampleSizeCalculator}) 
to form the \textsc{Unseen} group.

Eventually, we obtain the following two groups of code snippets:
\begin{itemize}
    \item \textsc{Unseen}: 10,000 Python function-level code snippets that \Code{WizardCoder} has never been exposed to.
    \item \textsc{Accessed}: 10,000 Python function-level code snippets from the training set of \Code{WizardCoder}.
   
    % \item \textsc{Unseen}: 1,000 Python functional code snippets that WizardCoder has never been exposed to.
\end{itemize}

\begin{table}[t]

\footnotesize
\centering
\setlength{\tabcolsep}{1.1mm}
%\vspace{-0.3cm}
\caption{Statistics of features from function-level code snippets in the two groups and the results of Mann-Whitney U test and Cliff’s Delta effect size test.}
\vspace{-0.2cm}
\label{tab:characteristc}
\begin{tabular}{l r r r r r r}
\toprule
& &\#prompt\_lines&\#body\_lines&complexity & \#comments\\

\midrule
 \multirow{4}{*}{\rotatebox{90}{\textsc{Unseen}}}&Min& 2 & 7& 1 & 0 \\
&Median&23 & 17 & 4 & 1\\
&Mean&27.9 & 28.8 & 6.5 & 3.3\\
&Max& 119 & 2467 & 317 & 244\\
\midrule
\multirow{4}{*}{\rotatebox{90}{\textsc{Accessed}}} &Min& 2 & 7 & 1 & 0 \\
&Median& 21 & 16  & 4 & 1\\
  &Mean& 25.6 & 24.6 & 5.4  & 2.8 \\
  & Max & 188 & 1583 & 751 & 326\\
\midrule
& $p$-value & $<$0.01 & $<$0.01 & $<$0.01 & $<$0.01 \\
\midrule
& Cliff’s $\delta$ & -0.07 & -0.07 & -0.10 & -0.03 \\
& Eff. Level & Negligible & Negligible & Negligible & Negligible\\
\bottomrule
\end{tabular}
\vspace{-0.5cm}
\end{table}

Table~\ref{tab:characteristc} presents some feature statistics of these samples in two different groups, including the number of prompt lines, %you need to explain your prompt process somewhere, right now you just say "complete the code snippet" that is unclear to many people how you conduct the experiment
%% I move the experiment setup to the front
the number of function body lines, the cyclomatic complexity~\cite{mccabe1976complexity} of the function, and the number of comments in function body. We compare these features between two groups using Mann-Whitney U Test~\cite{mcknight2010mann} and Cliff’s Delta effect size test~\cite{cliff1993dominance}. The Mann-Whitney U test determines whether there is a significant difference in their median values~\cite{gao2024characterizing,xiao2022recommending}, while the Cliff’s Deltaquantifies the extent of this difference~\cite{xu2021multi}.
Our analysis reveals that although the $p$-value is less than 0.01, %indicating a significant difference in the medians of the two groups,
the effect size is negligible. 
%This suggests that the practical difference between the groups is minimal despite the statistical significance. Therefore, we can infer that when the model completes the code snippets in these two groups, the key differentiation is whether it is independently creating content during the task, i.e., whether it has had prior exposure to the snippets or not, rather than inherent differences in the code characteristics between the groups.
%not sure i got your point here. it looks to me you can only say "there is no inherent differences in the code characteristics between the groups". 
%% yes, it's my point！ 
This suggests that there are no substantial inherent differences in the code characteristics between the two groups.
% However, we sort all potential samples in descending order based on copies number, and select an equal number of samples as \textsc{Accessed} group, specifically 385, to form the \textsc{Unseen} group. The reason for choosing based on descending copies is that these commonly occurring GPL-licensed files originate from very renowned GPL-licensed projects. The attribution of these codes is rarely disputed, and the legal risk associated with plagiarizing these codes is higher, given that the open-source community is well aware of their GPL licensing.

% It's worth noting that in the process of constructing the \textsc{Unseen} group, we deliberately exclude samples that, despite being under restrictive licenses, bear resemblance to the code present in the training set. Our findings revealed that xxxx (x\%) of the functions in the training set (involving xxxx (xxx\%) of the files) are similar to code snippets under restrictive licenses. This observation underscores the limitations of the current license filtering mechanisms of training sets for LLMs, which primarily rely on filtering at the project license level and struggle to completely exclude GPL. The potential reason behind this discrepancy, as unveiled by existing research~\cite{wolter2023open}, could be the inconsistency between the license of the code repository and the licenses of the individual files. 

\subsubsection{Features to characterize String Similarity} \label{ss:standardofsimilarity}
Considering that copyright law only protects expression rather than ideas~\cite{Copyrightfaq, autry2002toward}, we employ fundamental text similarity metrics including BLEU-4~\cite{papineni2002bleu}, Jaccard similarity based on MinHash~\cite{li2023starcoder}, and similarity based on edit distance~\cite{levenshtein1966binary,dong2024generalization} to measure the similarity between the function bodies output by the model and the original implementations in both \textsc{Accessed} and \textsc{Unseen} groups.

Inspired by judicial considerations of \textit{striking similarity} in fields like music~\cite{autry2002toward} and code's unique traits, we incorporate additional literal features that capture both structural and stylistic elements to characterize \textit{striking similarity}, including the number of function body lines, cyclomatic complexity, and comment similarity. The rationale behind this is as follows:
\begin{itemize}
    \item Number of function body lines.
Given the inherent nature of LLMs in generating code through token-by-token prediction~\cite{yang2024unveiling}, it implies that in scenarios of independent creation, longer code lengths reduce the likelihood of achieving occasional similarity.
    \item Cyclomatic complexity~\cite{mccabe1976complexity}. Determined by decision points, it increases potential paths through a function. More decision points expand the possibility space, significantly reducing the chance of occasional similarity.
    \item  The similarity of comments. 
Different developers have unique habits when it comes to writing comments. Unlike code, natural language possesses a higher degree of flexibility~\cite{allamanis2016convolutional}, making the similarity in comments within code exceedingly rare. %any reference would be more compelling %%done
\end{itemize}
\begin{figure}[H]
    \centering
\vspace{-3mm}
    \includegraphics[width=\linewidth]{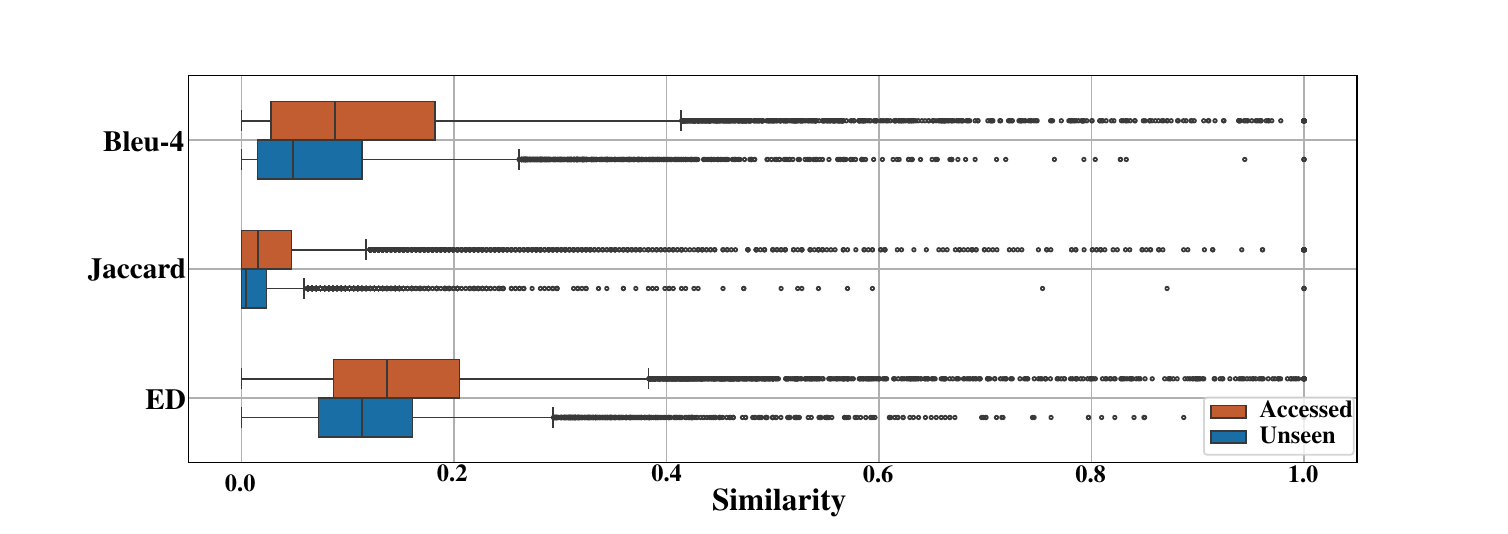}

    \caption{The similarity between output of WizardCoder and the corresponding open-source code in two groups.}
\vspace{-4mm}
    \label{fig:distribution}
\end{figure}

\begin{figure*}[t]
    \centering
    \subfigure[]{\includegraphics[width=0.32\linewidth]{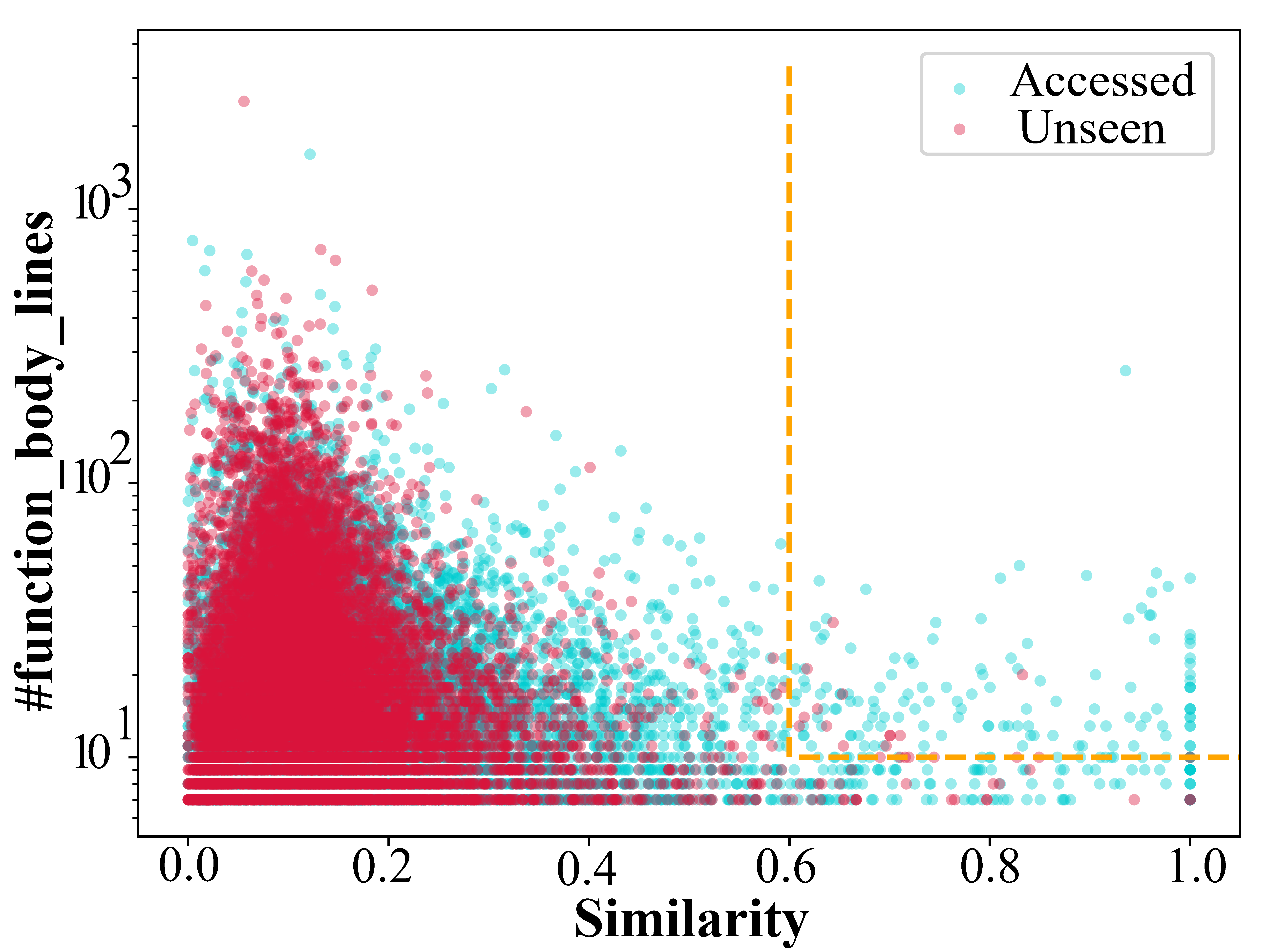}\label{fig:bodyline}}
    \subfigure[]{\includegraphics[width=0.32\linewidth]{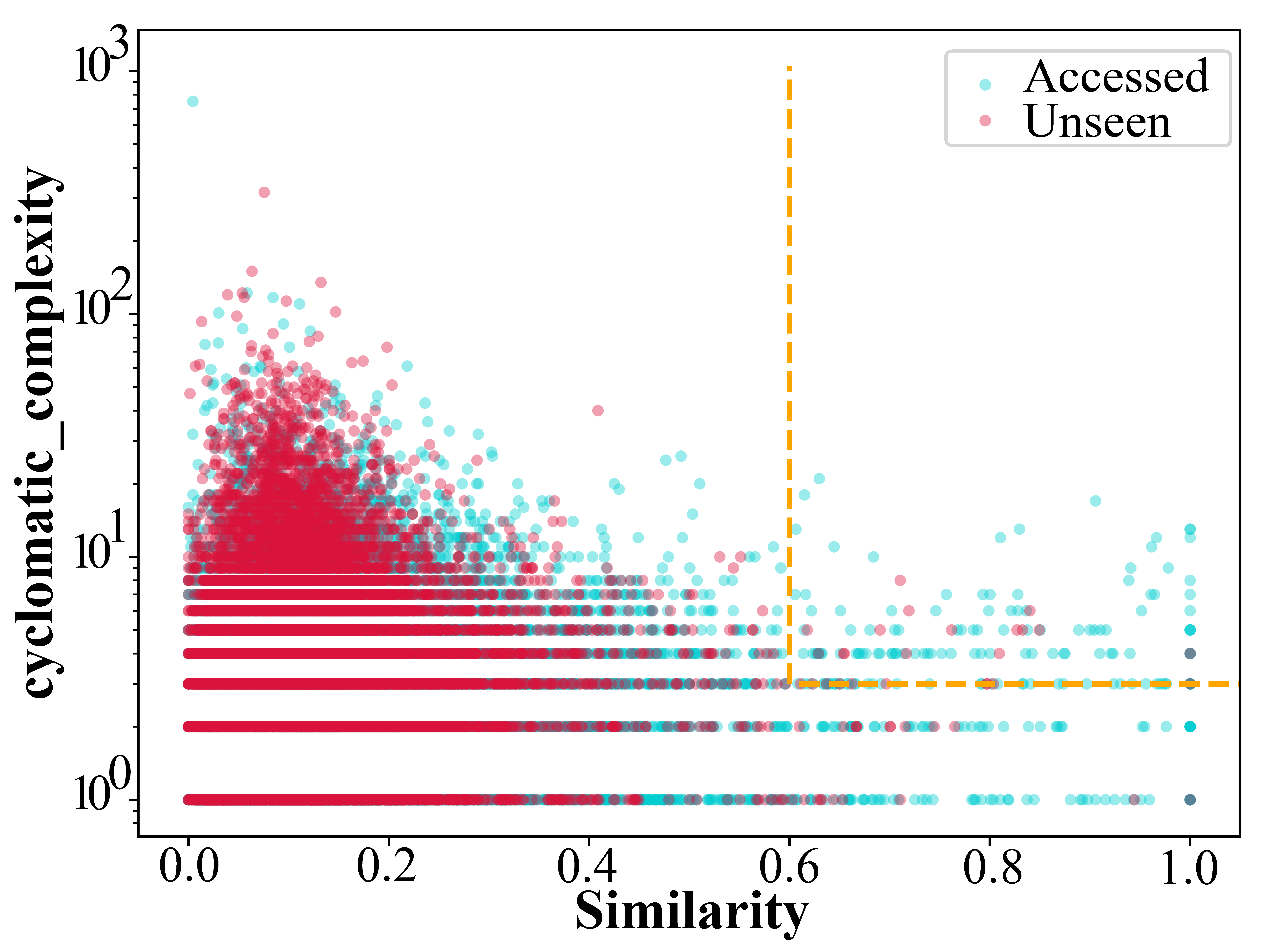}\label{fig:cc}}
    \subfigure[]{\includegraphics[width=0.32\linewidth]{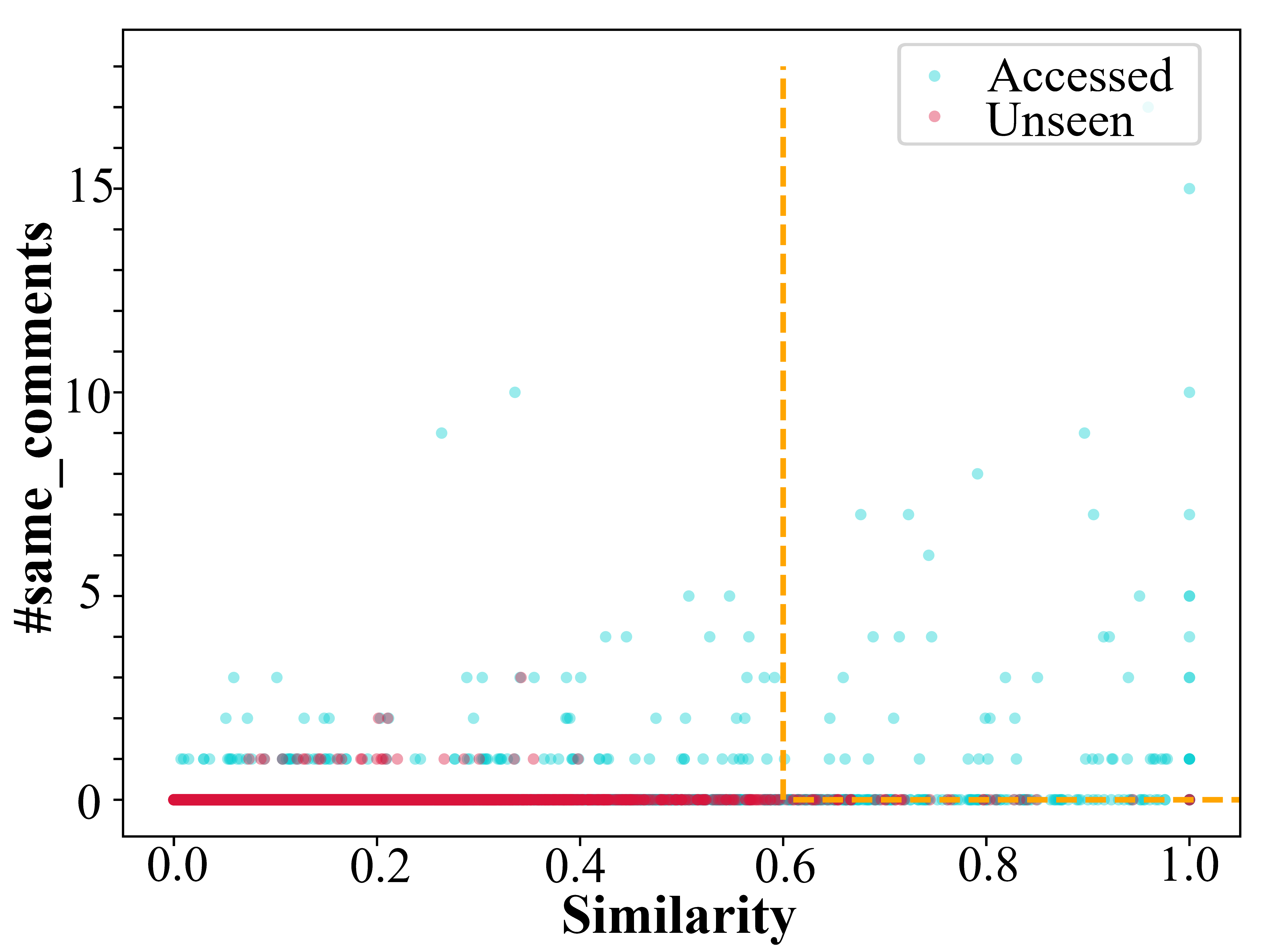}\label{fig:cmt}}
    \vspace{-2mm}
    \caption{The distribution of similarity between the generated code and the corresponding open-source implementations in two groups, in relation to the number of function body lines, cyclomatic complexity, and the number of same comments. The similarity value is the maximum of the three text similarity metrics.} 
    \label{fig:sim_feature}
    \vspace{-2mm}
\end{figure*}

\subsection{Results}

Figure~\ref{fig:distribution} shows the similarity between the outputs generated by \Code{WizardCoder} and the corresponding open-source %implementations 
code
in two groups. When \Code{WizardCoder} completes code for \textsc{Unseen} group, which simulates scenarios of independent creation, it generally produces code with lower similarity to the corresponding open-source code compared to its output for the  \textsc{Accessed} group, across all similarity metrics.
However, we observe some \textsc{Unseen} cases with exceptionally high similarity, occasionally reaching a score of 1. These findings suggest that while text similarity metrics can provide indication of differences in generated code between the two scenarios, they are insufficient on their own to definitively determine \textit{striking similarity} and, consequently, non-independent creation.

Figure~\ref{fig:bodyline} illustrates the distribution of similarity between the generated code and the corresponding open-source code in two groups, in relation to the number of function body lines of the open-source code. We observe that only for \textsc{Accessed} group does the LLM frequently generate highly similar code snippets (similarity $>$ 0.6) for functions with longer body lengths ($>$ 10 lines). In contrast, for longer functions from \textsc{Unseen} group, the similarity scores are generally lower. As shown in Figure~\ref{fig:cc}, a similar phenomenon is observed with cyclomatic complexity. Meanwhile, Figure~\ref{fig:cmt} reveals a stark contrast in generated comments between the two groups.  For \textsc{Unseen} group, the LLM rarely generates comments identical to those in the open-source code. However, for the \textsc{Accessed} group, it frequently produces comments that match those in the open-source code. In extreme cases, %the number of identical comment sentences can exceed 15.
the identical comments can exceed 15 sentences.

These observations reveal distinct patterns in code generation between \textsc{Accessed} and \textsc{Unseen} groups. Such distinctions suggest that certain combinations of these features might effectively indicate non-independent creation. To formalize this insight, we seek to establish a quantitative standard for identifying instances of \textit{striking similarity}.
Based on our analysis of \Code{WizardCoder}'s code generation results in these two simulated scenarios, we establish an initial standard for \textit{striking similarity} between generated code and the open-source code:
\begin{itemize}

    \item number of function body lines $>$ 10
    \item cyclomatic complexity $>$ 3
    \item text similarity (maximum of the three metrics) $>$ 0.6
    \item number of identical comments $>$ 0
\end{itemize}
The first two criteria are attributes of the open-source code snippet, while the latter two describe the relationship between the generated code and the corresponding open-source code snippet.
Under this standard, we identify 24 instances, all from \textsc{Accessed} group, i.e., the non-independent creation scenario. In the simulated independent creation scenario (\textsc{Unseen} group), none of the 10,000 functions generated by \Code{WizardCoder} meets this standard.

\begin{figure*}[t]
    \centering
    \includegraphics[width=\linewidth]{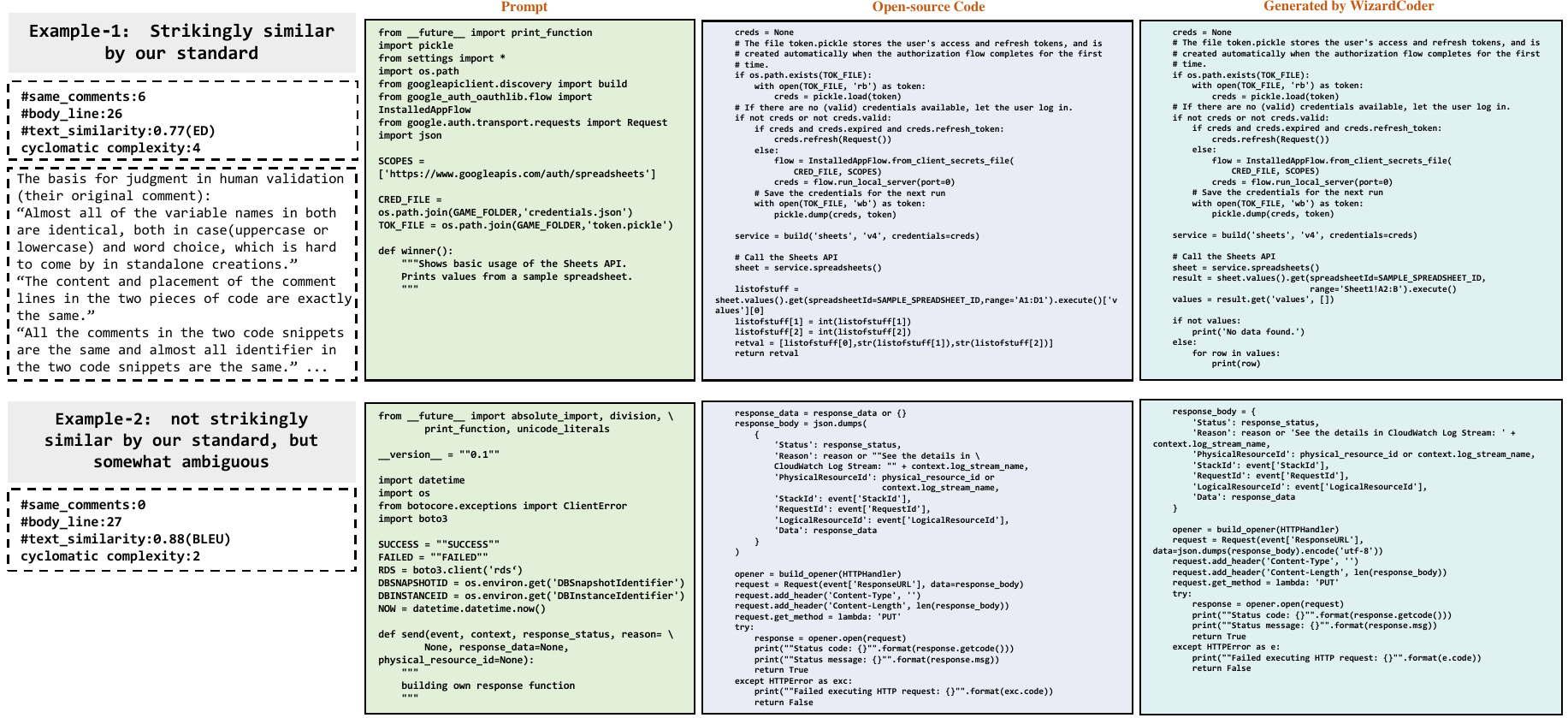}
\vspace{-4mm}
    \caption{Examples of cases above and below our striking similarity standard.}
\vspace{-4mm}
    \label{fig:ss_example}
\end{figure*}

\subsection{Validation}
Through the above experiments, we establish a preliminary standard for \textit{striking similarity} between LLM outputs and open-source code. 
This standard, when met, suggests a potential copying relationship rather than independent creation. 
%To ensure this standard reasonably reflects potential legal risks and remains applicable across various LLMs, 
%we conduct validation of this standard.
To validate the preliminary standard, we employ a three-step process: constructing new code samples, expanding our analysis to additional LLMs, and conducting an expert evaluation to assess the standard's validity and applicability.

 % First, we create two additional groups of code snippets to observe the output of two LLMs: WizardCoder and Poro (we choose Poro because ....). The results confirm the applicability of the standard. 
 %at the beginning try to be clear about your following logic flow

First, using the same methods employed in constructing \textsc{Accessed} and \textsc{Unseen} groups, we create two additional groups: \textsc{Accessed\_Eval} and \textsc{Unseen\_Eval}, each containing 10,000 samples that do not overlap with the original \textsc{Accessed} and \textsc{Unseen} groups. We then utilize \Code{WizardCoder} and \Code{Poro-34B-chat} to complete the functions for these 20,000 samples. We choose \Code{Poro} because it is a general model, yet shares the same code-related training data with \Code{WizardCoder}. This similarity in training data allows us to use the same two groups of samples we constructed, while also testing the standard's applicability to a more general model.

Among the outputs generated by these two models, 31 from \Code{WizardCoder} and 2 from \Code{Poro} meet our standard of \textit{striking similarity}. %you may create a phrase template for striking similarity
All 33 samples are from the \textsc{Accessed\_Eval} group, yielding a precision of 100\%. 

To further validate the standard, we assemble a diverse panel: five developers with over six years of coding experience and three lawyers specializing in software intellectual property. %%note, all interations with people may require ethical statement by the reviewers
This composition ensures both technical and legal perspectives in evaluating these 33 samples. The eight reviewers are tasked with determining whether independent creation can be ruled out based solely on the comparison between the code pairs, without knowledge of their origins. This approach ensures the standard's applicability to outputs from more advanced models, as the evaluation focuses purely on code characteristics. 
On average, each reviewer identifies 32 out of the 33 sample pairs as cases where independent creation can be excluded, implying a potential coping relationship. 

The experts cited three categories of reasons for their judgments:
\begin{itemize}
    \item Textual and structural similarities: This includes identical or similar variable names, overall textual resemblance, and similarities in code structure (such as indentation and blank lines).
    \item Logical and functional similarities: This encompasses similar approaches to problem-solving, comparable use of specific programming constructs, and similarities in core logic.
    \item Comment and unique feature similarities: This involves comments similarity (including content and formatting), as well as any distinctive elements like special punctuation or unusual coding patterns.
\end{itemize}

An example of \textit{striking similarity} is shown in Figure~\ref{fig:ss_example} (Example-1). In this example, the striking similarities are evident in multiple aspects: identical variable names (both in case and word choice), matching comment content and placement, and highly similar code structure. As noted by experts, such extensive similarities would be ``hard to come by in standalone creations." We generally find that the characteristics the experts are looking for align well with our \textit{striking similarity} standard.

Based on these results, we believe that our proposed standard serves as a reasonable preliminary standard for identifying instances with \textit{striking similarity }that may indicate potential legal risks. While not intended to establish definitive legal guidelines, this standard offers insights for exploring the complex landscape of AI-generated code and associated copyright concerns. 
%Based on this standard, we construct a license compliance evaluation framework and benchmark for evaluating existing LLMs.% (in the following section?).

\begin{small}
\begin{summary-rq}
\vspace{-1mm}
\textbf{Summary:}

% The reasonable standard of \emph{striking similarity} in LLM-generated code lies in a combination of specific thresholds across multiple metrics. Our study indicates that striking similarity occurs when the generated code exhibits: (1) high text similarity ( $>$ 0.6), (2) function body length exceeding 10 lines, (3) cyclomatic complexity greater than 3, and (4) more than one identical comment. This standard, derived from analyzing WizardCoder's outputs across \textsc{Accessed} and \textsc{Unseen} groups, effectively identifies instances highly unlikely to result from independent creation, thus providing a basis for evaluating LLMs' license compliance in code generation.

Through a comparative analysis of LLM-generated outputs derived from previously accessed and unseen code samples, we establish and validate a preliminary standard of \textit{striking similarity} that effectively excludes the possibility of independent creation.

%This standard is met when the generated code exhibits:  (1) function length exceeding 10 lines, (2) cyclomatic complexity greater than 3, (3) text similarity above 0.6, and (4) at least one identical comment.

% The reasonable standard of "striking similarity" in the context of code generation by LLMs can be situated at a point where the identified similarities between code snippets are substantial to the extent that they are unlikely to have occurred independently. This standard could encompass a combination of text similarity metrics and additional features such as function length, cyclomatic complexity, and comment similarity. By establishing a threshold where the similarities surpass what could reasonably be attributed to chance or independent creation, researchers can detect instances of potential concern, providing a basis for further evaluation of LLMs' compliance with legal and ethical standards in code generation tasks.
\vspace{-4pt}
\noindent\rule{\textwidth}{0.8pt}

\textbf{Implications:}
(1) Text similarity alone cannot determine non-independent creation in LLM-generated code. LLMs can produce highly similar code even for unseen simple functions. This necessitates using adequately complex code, as defined by our established standard,  when constructing LLM evaluation benchmarks.
(2) LLMs can memorize and reproduce comments from training data, guiding our development of evaluation frameworks for compliance capabilities. This finding informs our subsequent investigation into LLMs' ability to recall and generate license information in file headers.
\vspace{-3pt}
\end{summary-rq}
\end{small}

%(1) Text similarity metrics alone are insufficient to determine whether LLM-generated code is independently created. We found that even for unseen code samples, LLMs can produce highly similar code when the functions are simple. This observation underscores the necessity of using code with adequate complexity, as defined by our established standard, when constructing benchmarks to assess LLMs. (2) Our experiments demonstrate that LLMs possess the ability to memorize and reproduce comments from their training data. This finding provides valuable direction for developing evaluation framework and benchmark to evaluate LLMs' compliance capabilities.

\begin{figure*}[t]
    \centering
    \includegraphics[width=\linewidth]{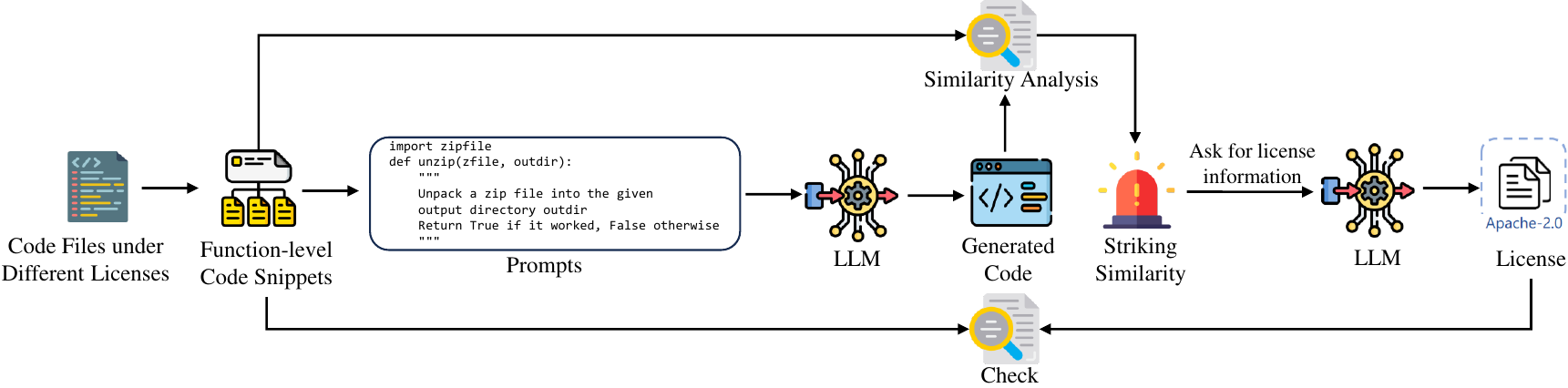}
\vspace{-4mm}
    \caption{Overview of the evaluation framework.}
\vspace{-2mm}
    \label{fig:framework}
\end{figure*}
\section{Evaluation Framework and Benchmark for LLM License Compliance}
\subsection{Evaluation Framework}
As illustrated in Figure~\ref{fig:framework}, we propose a framework to evaluate LLMs on license compliance in code generation. This framework is grounded in the empirical findings that LLMs, in non-independent creation scenarios, may generate code strikingly similar to existing implementations, accompanied by identical comments. This observation leads to a crucial question:  If LLMs can reproduce code and comments with high fidelity, can they also accurately output the associated license information typically found in file comments? Our framework builds upon this insight, positing that when an LLM generates code strikingly similar to existing code, it should also be capable of providing the corresponding license or copyright information. This principle forms the foundation of our evaluation methodology, linking the generation of strikingly similar code to the ethical and legal responsibility of proper attribution and license compliance.

The framework operates by first constructing a benchmark comprising functions from widely reused code files that have explicit copyright information in their file header comments. The detailed construction method is elaborated in Section\ref{benchmark}. The structure of the functions in this benchmark aligns with that as shown in Figure~\ref{fig:codeexample}. Our prompt consists of the first four components, and we task the LLM to complete the function body. Subsequently, we conduct a similarity analysis between the LLM's output and the corresponding open-source code snippet. If the output meets our established standard for \textit{striking similarity}, we prompt the LLM, in the form of a follow-up inquiry, to output the license information. Finally, we compare the license information provided by the LLM with the actual license of the open-source code snippet.

\subsection{Benchmark~\BenchmarkName}
\label{benchmark}

\subsubsection{Construction Method}
We construct our benchmark named \BenchmarkName~by mining code files from the open-source ecosystem that have explicit license information and are widely reused. This process adheres to the standards we previously established. The specific steps are as follows.

\textbf{Data Source---World of Code.} We utilize World of Code (WoC)~\cite{ma2019world} as the source for constructing the benchmark. WoC is a comprehensive infrastructure for mining version control data across the entire open-source software ecosystem. It aggregates Git objects, including commits, trees, and blobs~\cite{Gitobjects} on platforms like GitHub, Bitbucket, and GitLab. WoC provides several key-value databases that enable efficient querying of relationships between different entities~\cite{gao2024pyradar,gao2022variability}. For instance, the blob-to-project (b2p) database maps each blob to all projects that contain it, enabling efficient tracking of code reuse across projects.
For our benchmark, we use version U of WoC, released in October 2021. This version encompasses over 173 million Git repositories, 3.1 billion commits, 12.5 billion trees, and 12.4 billion blobs~\cite{wocoverview}.

\textbf{Collecting Python Code Files.}
Our first step utilizes WoC's c2fbb database, which maps each commit to its corresponding commit file name, new blob (post-commit), and old blob (pre-commit). We filter these commits to focus on Python files based on file extensions. By selecting the new blobs associated with these commits and removing duplicates, we obtain a dataset of 700,867,958 unique Python file blobs.

\textbf{Collecting licensed function-level code snippets.}
We identify blobs containing explicit license information in their file header comments. Using the b2p database, we quantify each selected blob's occurrence across different projects. We then sort these license-containing blobs in descending order of project count, aiming to identify widely reused code files with clear license information. The presence of license information in file headers indicates clear copyright attribution, while widespread reuse suggests general acknowledgment and acceptance of this copyright information.

We iterate through the sorted blobs in descending order, segmenting each file into function-level code snippets. We then apply a filtering process to select snippets that not only adhere to the three principles described in Section~\ref{sec:codesamples} but also meet the preconditions of our \textit{striking similarity} standard (function body $>$ 10 lines,  complexity $>$ 3, and  comment number $>$ 0). From this filtered set, we select the top 10,000 qualifying code snippets as candidate samples.
We further refine these 10,000 candidate samples removing code snippets that are explicitly indicated as copied or derived from other sources (identified by keywords like ``copied from" or ``taken from") and excluding code snippets with dual licenses to ensure license clarity. We also perform deduplication based on function signatures and docstrings, retaining only one instance where these elements are identical. This rigorous filtering process ultimately yields 4,187 unique function-level code snippets for our \BenchmarkName~benchmark.

\begin{figure}[b]
    \centering
    \vspace{-3mm}
    \includegraphics[width=\linewidth]{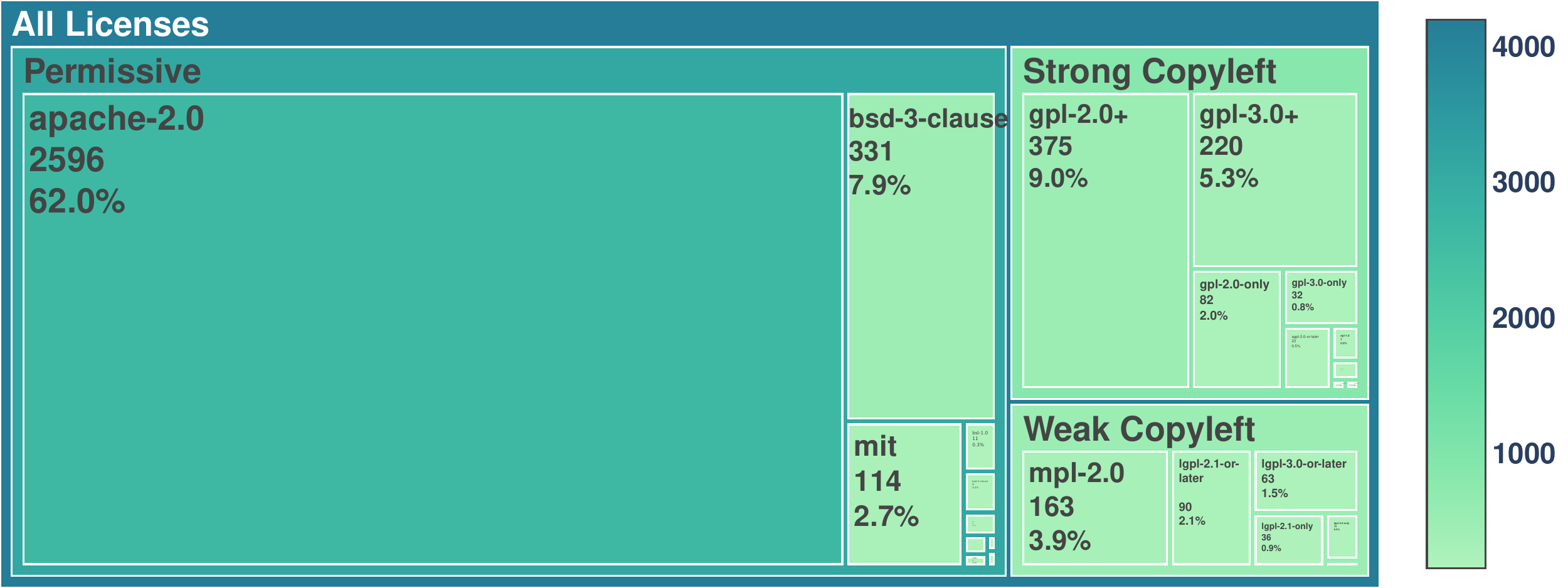}
\vspace{-5mm}
    \caption{The distribution of licenses in the benchmark.}
\vspace{-2mm}
    \label{fig:license_distribution}
\end{figure}
\subsubsection{Benchmark Characteristics}
The detailed characteristics of \BenchmarkName~are as follows.

\textbf{License Information Accuracy.} %In constructing \BenchmarkName, 
We employ a keyword and rule-based method proposed by Xu et al.~\cite{xu2023licensecom} to identify licenses. 
To evaluate the accuracy of license information in \BenchmarkName, we randomly sample 352 samples (95\% confidence level, 5\% confidence interval~\cite{SampleSizeCalculator}). We manually review the file header comments and function docstrings of these samples to verify the correctness of the license information extracted from the file headers and to check for any exception statements in the docstrings.
%Upon examination, 
We find that the license information for all 352 samples is correct. %Based on this result, 
Therefore, we believe that the license information in \BenchmarkName is highly reliable and can serve as a robust foundation for evaluating the compliance capabilities of LLMs.

\textbf{License Distribution.}
Licenses can be categorized into three types based on their level of permissiveness: Permissive, Weak Copyleft, and Strong Copyleft~\cite{xu2023licensecom}. Copyleft licenses mandate that software which modifies or utilizes existing software must be licensed under the same terms, unless explicitly specified otherwise (e.g., \textit{GPL-3.0}). 
% Non-compliance with such licenses has led to numerous legal disputes~\cite{huang2023detecting}. 
It is crucial to note that permissive licenses also require adherence to their terms. A license is essentially a conditional authorization~\cite{Meeker2020OpenSource}, and failure to comply with conditions stipulated in permissive licenses can also result in non-compliance issues~\cite{huang2023detecting}, e.g., Apache-2.0 and CC-BY-4.0 licenses require users to provide a statement of changes made to the original software.

Therefore, in \BenchmarkName, we retain a natural distribution of licenses, covering the three different types of licenses. As shown in Figure~\ref{fig:license_distribution}, out of 4,187 function-level code snippets, the number of licenses for Permissive, Weak Copyleft, and Strong Copyleft are 3,073 (73.4\%), 369 (8.8\%), and 745 (17.8\%), respectively. Among permissive licenses, \textit{Apache-2.0} is the most prevalent, accounting for 2,596 snippets (62.0\% of the total). For copyleft licenses, the most common are \textit{GPL-2.0-or-later} with 375 snippets (9.0\%), \textit{GPL-3.0-or-later} with 220 snippets (5.25\%), and \textit{MPL-2.0} with 163 snippets (3.9\%).
\begin{table}[t]
\centering
\footnotesize
\caption{Statistics of \BenchmarkName.}
 \setlength{\tabcolsep}{1.2mm}
  \label{tab: codeprofile}
  \renewcommand{\arraystretch}{1.2}
  \begin{threeparttable}
  \begin{tabular}{lrrrrr}
    \toprule
    Metric & Min & Median & Mean & Max &Distribution\tnote{$*$}\\
    \midrule
    \#prompt\_lines & 2 & 25 & 28.5 & 119 &\adjustimage{height=0.3cm,valign=m}{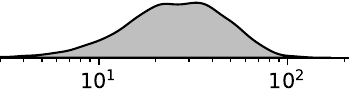}\\ 
    \#prompt\_tokens & 25 & 265 & 310.3 & 964 & \adjustimage{height=0.3cm,valign=m}{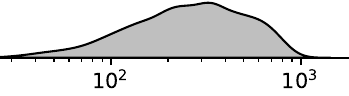}\\ 
    \#body\_lines & 11 & 28 & 36.8 &  398 &\adjustimage{height=0.3cm,valign=m}{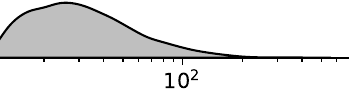}\\
   \#body\_tokens & 64 & 356 & 479.8 & 5418 &\adjustimage{height=0.3cm,valign=m}{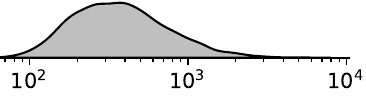} \\ 
   \#project\_reuse & 16 & 54 & 245.0 & 42,476 &\adjustimage{height=0.3cm,valign=m}{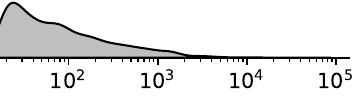}\\ 
    \#comments & 1 & 4 & 5.7 & 159 &\adjustimage{height=0.3cm,valign=m}{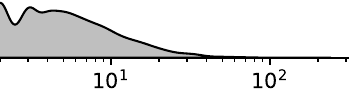}\\ 
     cyclomatic\_complexity & 4 & 7 &9.1 & 95 &\adjustimage{height=0.3cm,valign=m}{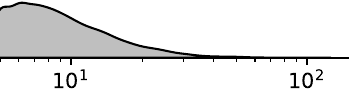}\\ 
   
    \bottomrule
    \end{tabular}
    
    \begin{tablenotes}
    \footnotesize
    \item[$*$] We increment all values by one to plot the distribution in log-scale.
    %For better clarity, we add 1 to the overall value
    \end{tablenotes}
    \end{threeparttable}
    \label{tab:codemetric}
    \vspace{-5mm}
\end{table}   

\begin{table*}
\centering
\footnotesize
    \caption{Performance of 14 LLMs on \BenchmarkName. \checkmark means publicly available weights and $\times$ means unavailable weights.}
        
\setlength{\tabcolsep}{1.2mm}
  \renewcommand{\arraystretch}{1.2}
  \begin{tabular}{llcc|cc|cc|cc|c}
    \toprule
    &Model  & HumanEval &Weights&  \#striking\_sim & $Acc$ &\#permissive& $Acc_{p}$ &\#copyleft & $Acc_{c}$ & \textsc{LiCo} \\
    \midrule
    \multirow{8}{*}{\makecell{General\\LLM}}&GPT-3.5-Turbo\cite{achiam2023gpt}&  72.6&$\times$& 29 (0.69\%) & 0.72 &26 &0.81 &3 &0.0 & 0.373\\ 
    &GPT-4-Turbo\cite{achiam2023gpt}& 85.4 &$\times$& 25 (0.60\%) & 0.72 &22 &0.82 & 3 & 0.0 & 0.376\\ 
    &GPT-4o~\cite{achiam2023gpt}& 90.2 &$\times$& 47 (1.12\%)&  0.74 &41 &0.85 &6 & 0.0 & 0.385\\
    &Gemini-1.5-Pro\cite{reid2024gemini}& 71.9 &$\times$& 41 (0.98\%)&  0.59 &39 &0.62 & 2 &0.0 & 0.317\\
    &Claude-3.5-Sonnet\cite{claude-3-5-sonnet} & 92.0&$\times$& 84 (2.01\%) &  0.69 &79& 0.71 &5 & 0.4 &0.571 \\
    &Qwen2-7B-Instruct\cite{Qwen2}& 79.9 &\checkmark& 20 (0.48\%) &  0.95 &20 & 0.95 &0& - &0.985 \\
    &GLM-4-9B-Chat\cite{team2024chatglm}& 71.8&\checkmark& 0 (0.0\%) &  - &- &- & - & - &1.0\\
    &Llama-3-8B-Instruct\cite{llama3modelcard}&  62.2&\checkmark& 1 (0.02\%) &  0.0 &1 & 0.0 & 0& - & 0.714 \\
    \midrule
    \multirow{6}{*}{\makecell{Code\\ LLM}}&DeepSeek-Coder-V2\cite{zhu2024deepseek}  & 90.2 &\checkmark& 37 (0.88\%) & 0.0  & 36 & 0.0 &1 &0.0 & 0.142\\ 
    &CodeQwen1.5-7B-Chat\cite{Qwen1.5}& 83.5 &\checkmark& 17 (0.41\%) & 0.24 & 17 & 0.24 & 0 & - & 0.781\\ 
   &StarCoder2-15B-Instruct\cite{lozhkov2024starcoder}& 72.6 &\checkmark &13 (0.31\%) & 0.23 & 13 &0.23 & 0 & -&  0.780 \\ 
   & Codestral-22B-v0.1\cite{Codestral}& 61.5 &\checkmark & 91 (2.17\%) & 0.73 &87 &0.77&4 &0.0 & 0.360\\ 
    &CodeGemma-7B-IT\cite{team2024codegemma}& 56.1 &\checkmark& 3 (0.07\%) & 0.33 &3 & 0.33 & 0 & - & 0.809\\ 
    &WizardCoder-Python-13B\cite{luo2023wizardcoder}  & 64.0 &\checkmark& 27 (0.64\%)  & 0.04 &26&0.04 &1 &0.0 & 0.153\\
    \bottomrule
    \end{tabular}

    \label{tab:acc}
    \vspace{-2mm}
\end{table*}
\vspace{-4mm}
\textbf{Code Metrics.}
Table~\ref{tab:codemetric} presents statistics of \BenchmarkName, including the number of lines in function bodies (\#body\_lines), cyclomatic complexity, the number of comments within function bodies (\#comments), and other relevant metrics. Due to our application of preconditions for striking similarity during the selection process, all functions in the benchmark satisfy the criteria of \#body\_lines $>$ 10, \#comments $>$ 0, and cyclomatic\_complexity $>$ 3.
The metric \#reuse\_projects indicates the number of projects that have reused the blob containing the function. A higher value suggests more widespread adoption in the open-source ecosystem.

As evident from Table~\ref{tab:codemetric}, all function-level code snippets in \BenchmarkName~generally exhibit high complexity and are extensively reused. This indicates that these code snippets are non-trivial while also being widely acknowledged by the open-source community in terms of their copyright information.
%(because the file header comments of the blobs where these functions are located contain explicit licensing information). 

\section{Evaluating LLMs on License Compliance}
\subsection{Experiment setup}
We evaluate 14 popular LLMs that 
%we exhaustively obtain (do we have evidence that they are representative??) and 
exhibit strong performance in code generation tasks (Pass@1 $>$ 0.5 on HumanEval) using \BenchmarkName. 
Table~\ref{tab:acc} presents the complete list of evaluated LLMs. Throughout the evaluation process, we consistently employ a one-shot approach using greedy decoding (temperature $=$ 0).

Effective assessment of license compliance requires a balanced consideration of two critical factors: a model's propensity to generate code strikingly similar to existing implementations, and its accuracy in providing licenses, with particular emphasis on copyleft licenses. This dual focus is essential, as it addresses both the risk of unintended code replication and the model's cognizance of licensing obligations.
To quantify this balance, we introduce a novel metric, 
%%we need justification/rational for this "NOVEL" metric, otherwise it comes out of nowhere and it's unclear if it's reasonable
\textsc{LiCo} (\underline{Li}cense \underline{Co}mpliance), calculated as follows:
\begin{equation}
\textsc{LiCo} = \frac{w_1 \cdot (1 - N)+ w_2 \cdot Acc_{p} + w_3 \cdot Acc_{c}}{w_1 + w_2 + w_3}
\end{equation}

Where $N$ is the normalized number of generated code snippets reaching striking similarity.  $Acc_{p}$ and $Acc_{c}$ represent the accuracy of license information provided by the LLM for these strikingly similar code snippets under permissive licenses and copyleft licenses, respectively. We set the weights as $w_1 = 1$, $w_2 = 2$, and $w_3 = 4$, emphasizing copyleft license compliance due to its associated legal risks. For missing accuracy metrics, we use a value of 1, assuming optimal performance in the absence of data. The \textsc{LiCo} score ranges from 0 to 1, with higher scores indicating less compliance risks.
It's crucial to note that a high \textsc{LiCo} score, particularly a score of 1 in the absence of any strikingly similar cases, is not meaningful if the model's code generation performance is poor. 
%As previously mentioned, compliance is only meaningful when the generated code is functionally correct. 
Models producing erroneous or chaotic code may naturally avoid striking similarities, resulting in high \textsc{LiCo} scores that lack practical significance.

%do not reflect true compliance capabilities. Therefore, the \textsc{LiCo} metric should always be interpreted in conjunction with the model's overall code generation performance metrics, such as HumanEval scores.

\subsection{Results}
\label{sec:results}
Table~\ref{tab:acc} presents the performance of 14 LLMs on \BenchmarkName. 
We first observe that the three LLMs currently performing best in code generation (\Code{GPT-4o}, \Code{Claude-3.5-Sonnet}, and \Code{DeepSeek-Coder-V2}) show significant variations in their results. They produce 47 (1.12\%), 84 (2.01\%), and 37 (0.88\%) strikingly similar cases, respectively, which are not insignificant proportions, indicating that compliance issues are not uncommon even among top-performing models.
Regarding the accuracy of providing license information, \Code{DeepSeek-Coder-V2} performs the poorest, unable to provide any license information, while \Code{GPT-4o} performs the best with an accuracy of 0.74. For the higher-risk copyleft licenses, only \Code{Claude-3.5-Sonnet} demonstrates good performance, which contributes to its highest \textsc{LiCo} score.
Among other models, \Code{GLM-4-9B-Chat} stands out by not producing any strikingly similar cases, and \Code{Qwen2-7B-Instruct} also demonstrates excellent compliance performance, achieving a license accuracy of 0.95 for its 20 strikingly similar cases, resulting in a \textsc{LiCo} score of 0.985. Furthermore, \Code{Codestral-22B-v0.1} generates the highest number of strikingly similar cases at 91, but maintains a relatively good accuracy of 0.73 in providing correct license information. Notably, \Code{WizardCoder-Python-13B} exhibits poor compliance performance, with a \textsc{LiCo} score of 0.153, and is almost incapable of providing any correct license information.

For strikingly similar code snippets under higher-risk copyleft licenses, we find that all LLMs perform poorly. Only \Code{Claude-3.5-Sonnet} provides some copyleft license information, while other LLMs have an accuracy of zero. We speculate that some closed-source LLMs may have implemented post-processing steps to avoid acknowledging outputs derived from copyleft-licensed code snippets. It is worth noting that \Code{StarCoder2}, in constructing its training set \Code{The Stack v2}, employed file-level license detection to exclude copyleft-licensed files. This approach may be a significant factor in explaining why the number of strikingly similar cases for copyleft licenses is zero for this LLM. Moreover, we observe that among general LLMs, open-source LLMs demonstrate superior compliance performance compared to closed-source LLMs. This finding suggests a potential correlation between model transparency and license compliance capabilities.

\section{Discussion}
\subsection{Implications}
In this section, we discuss the implications of our results for LLM providers, LLM users, open-source communities, and legal professionals:
\subsubsection{LLM providers}
Our findings highlight the need for LLM providers to enhance their LLMs' license compliance capabilities. This involves several key areas:

\textbf{Data Cleaning and License Detection:} Our empirical study in Section~\ref{sec:codesamples} reveals that despite the provider's efforts to exclude code files from copyleft-licensed repositories during \Code{Starcoderdata}'s construction, copyleft-licensed code still persists. This is often due to discrepancies between file-level and repository-level license information~\cite{wolter2023open}. The success of \Code{Starcoder2}'s file-level, fine-grained license detection strategy, which resulted in no strikingly similar cases under copyleft licenses in our evaluation, underscores the importance of meticulous data cleaning processes.

\textbf{Enhancing License-Code Association:} Despite generating non-negligible proportion of strikingly similar code, demonstrating impressive memorization capabilities, many LLMs fail to provide correct license information. This suggests ineffective learning of code-license associations during training. LLM providers should implement more sophisticated preprocessing techniques to strengthen these associations.
Considering autoregressive nature of these LLMs, which predict subsequent content based on preceding information, one potential area for exploration could be the positioning of license information during training. Placing license information after the code snippets might potentially enhance LLMs' ability to associate specific code with its corresponding license, presenting a promising direction for future research.

\textbf{Addressing Copyleft Information Suppression:} In Section~\ref{sec:results}, we find that most models tend to avoid providing information about copyleft code usage, likely due to implemented output filters that suppress such information. This approach is problematic as it provides users with incomplete or potentially misleading information, exposing them to higher legal risks. Instead of suppressing copyleft-related outputs, providers should implement post-processing steps that ensure proper attribution and accurate license information are included with generated code, regardless of the license type.
\subsubsection{LLM users}
Our evaluation reveals that many LLMs exhibit poor compliance capabilities. Users, especially commercial entities, must be aware of the potential legal risks associated with AI-generated code. Before incorporating AI-assisted development into their workflows, users should carefully evaluate the compliance capabilities of LLMs. If opting to use LLMs, it is crucial to employ code review tools to verify license compliance of generated code, particularly for code potentially derived from copyleft-licensed sources.
%Additionally, using LLMs from providers that offer legal guarantees, such as OpenAI, is a viable option to consider. These measures can help users minimize legal risks while benefiting from the productivity gains offered by AI-assisted code generation.

\subsubsection{Open-source communities}
The difficulties LLMs face in accurately providing license information underscore the risk of open-source projects' intellectual property being infringed upon in AI-driven development. While not obligated to facilitate AI training, open-source projects might consider adopting more explicit license declarations to protect their own intellectual property rights effectively. This could include embedding license information in individual files and adopting more granular licensing practices. Furthermore, as AI-assisted coding becomes more prevalent, open-source communities may need to revise their practices and principles. This could include developing guidelines for incorporating and attributing AI-generated code in projects, and establishing clear policies on how their own code should be used in AI training and generation processes.

\subsubsection{Legal professionals}
We find even top-performing code generation LLMs produce a non-negligible proportion (0.88\% to 2.01\%) of strikingly similar cases. Their varying abilities to provide correct license information (with \Code{GPT-4o} achieving 0.74 accuracy and \Code{DeepSeek-Coder-V2} failing entirely) highlight the complexity of license compliance in AI-generated code.
Our study underscores the need for clearer legal frameworks addressing AI-generated code and potential copyright infringements. We demonstrate that it is feasible to characterize non-independent creation in LLM outputs using specific features, opening new avenues for legal analysis. 
%We suggest that developing standardized methods for determining \textit{striking similarity} in AI-generated code could significantly benefit future legal considerations.  
These insights could serve as a reference for establishing more concrete legal standards in this emerging field, aiding legal professionals in cases involving AI-generated code.

\subsection{Threats to validity}
% While our study provides valuable insights into the license compliance capabilities of LLMs, it is important to acknowledge several limitations.

\subsubsection{Internal Validity} Our \textit{striking similarity} standard focuses on precision, potentially overlooking cases where LLMs generate code derived from open-source code but fall below our threshold (e.g., Example-2 in Figure~\ref{fig:ss_example}). There is no precise legal standard to exclude independent creation and any automated approach would be inherently ambiguous and would not predict precise decisions in a legal context. To mitigate this threat, we opt for a ``minimum" standard that emphasizes precision and interpretability, identifying cases that are likely not independently created (e.g., Example-1). This minimum standard aligns well with the qualitative characteristics experts look for when the decide on \textit{striking similarities}. If any of those standards are not met, as in Example-2, it would be unclear whether the two snippets would have been independently created. We must acknowledge that our standard may perform poorly on recall, which is hard to provide evidence in terms of recall due to such inherent ambiguity. Even with such a minimum standard, we are still able to obtain concerning results from state-of-the-art LLMs as shown in Table~\ref{tab:acc}. Furthermore, \BenchmarkName, consisting of 4,187 samples (code snippets), is a constructed benchmark that may not fully represent the vast diversity of real-world code. This limits our ability to determine if LLMs generate code infringing on open-source software outside our benchmark. Despite this constraint, the substantial findings within our focused dataset indicate that the identified compliance issues are likely prevalent in wider contexts as well.

\subsubsection{External Validity} our study primarily focused on Python code. While results may vary for other programming languages with different licensing practices, we believe that our evaluation framework and the methodology for constructing the benchmark are general and can be easily extended to other programming languages. Furthermore, We only addressed function-level code completion, overlooking potentially more severe compliance issues at class or project levels. However, our approach provides a foundation for investigating these broader contexts.

\section{Conclusion and Data Availability}
In this paper, our main contribution is the development of \BenchmarkName, the first benchmark for assessing LLMs' license compliance capabilities. To construct this benchmark, we conduct an empirical study on \textit{striking similarity} in LLM-generated code, establishing a preliminary standard for this concept. Using \BenchmarkName, we perform a evaluation of 14 LLMs, revealing significant license compliance shortcomings.

Although this study is only a preliminary attempt, and much more work is needed beyond the basic requirements of copyright laws, 
we believe our work could provide valuable insights for improving license compliance in AI-assisted software development. 
%protecting open-source developers' IP rights and mitigating legal risks for LLM users
% We provide a replication package
% on Figshare~\cite{figshare}.
\BenchmarkName~can be accessible at \cite{LiCoEval}.

\section*{Acknowledgment}
This work is sponsored by the National Natural Science Foundation of China 62332001.

%We would like to express our gratitude to all the experts involved in validation. 
\bibliographystyle{IEEEtran}
\bibliography{ref}

\end{document}